\documentclass[runningheads]{llncs}

\usepackage{float}
\usepackage{comment}
\usepackage[utf8]{inputenc}
\usepackage{circuitikz}
\usepackage{multirow}
\usepackage{enumitem}
\usepackage{amsmath}
\usepackage{calc}

\usepackage[T1]{fontenc}
\usepackage{graphicx}
\usepackage[english]{babel}
\usepackage{csquotes}

\usepackage{booktabs}
\usepackage{tabularx}
\usepackage{array}
\usepackage{ragged2e}
\usepackage{listings}
\usepackage{color}
\usepackage{soul}
\lstdefinestyle{mystyle}{
    basicstyle=\ttfamily\footnotesize,
    breakatwhitespace=false,         
    breaklines=true,                 
    captionpos=b,                    
    keepspaces=true,                 
    numbers=left,                    
    numbersep=5pt,                  
    showspaces=false,                
    showstringspaces=false,
    showtabs=false,                  
    tabsize=2
}
\usepackage{tikz}
\usetikzlibrary{decorations.pathreplacing}

\usepackage{todonotes}

\usepackage{biblatex}
\usepackage{caption}
\usepackage{subcaption}

\begin{document}
\title{Runtime Verification Containers for Publish/Subscribe Networks}
%
%
\author{Ali Mehran\and Dogan Ulus}
%
%
\authorrunning{A. Mehran and D. Ulus}
%
\institute{Boğaziçi University, Istanbul, Türkiye}
%
\maketitle              
\begin{abstract}
Publish/subscribe (pub/sub) networks are a cornerstone of modern distributed systems, playing a crucial role in applications like the Internet of Things (IoT) and robotics. While runtime verification techniques seem ideal for ensuring the correctness of such highly dynamic and large-scale networks, integrating runtime monitors seamlessly into real-world industrial use cases presents significant challenges. This paper studies modern containerization technology to deploy runtime verification tools to monitor publish/subscribe networks with a performance focus. Runtime verification containers are lightweight and deployable alongside other containerized publisher and subscriber participants. Each runtime verification container monitors message flow, enabling runtime verification of network behavior. We comprehensively benchmark the container-based approach using several experiments and a real-world case study from the software-defined vehicle domain.

\end{abstract}

\section{Introduction}
\label{sec:intro}

Modern software applications have become increasingly distributed, with components on edge devices and in the cloud. These components interact with each other and external services asynchronously, requiring robust and reliable communication mechanisms. Leveraging a publisher/subscriber (pub/sub) network architecture is a prevalent approach to facilitate communication within these distributed systems. Pub/sub networks enable publisher participants to disseminate messages, often containing data or event notifications, to a pool of subscriber participants. These subscribers, in turn, have expressed interest in receiving specific categories of information by subscribing to relevant topics. Such loosely coupled communication patterns, while enabling scalability and flexibility, also introduce challenges in ensuring correct operation, maintaining software across iterations, and managing the complexity of distributed system development.

\begin{figure}
\begin{center}
\begin{tikzpicture}

\draw [draw=black, fill=white, very thick] (1,1) rectangle (3,3);
\draw [draw=black, fill=white, very thick] (1.2,0.8) rectangle (3.2,2.8);
\draw [draw=black, fill=white, very thick] (1.4,0.6) rectangle (3.4,2.6);
\coordinate[align=center,text width=1,label=center:{\parbox{2cm}{\centering Application\\Pub/Sub\\Participants}}] (R) at (2.4,1.6);

\draw [draw=black, fill=white, very thick] (6.6,0.6) rectangle (7.2,2.6);
\coordinate[align=center,text width=1,label=center:{\rotatebox[origin=c]{90}{\parbox{2cm}{Filter}}}] (R) at (6.9,2.1);

\draw [draw=black, fill=white, very thick] (7.4,0.6) rectangle (9.4,2.6);
\coordinate[align=center,text width=1,label=center:{\parbox{2cm}{\centering Runtime\\Verification\\Engine}}] (R) at (8.4,1.6);

\draw [draw=black, fill=white, very thick] (9.6,2.0) rectangle (11.6,2.6);
\coordinate[align=center,text width=1,label=center:{\parbox{2cm}{\centering Test DB}}] (R) at (10.6,2.3);

\draw [draw=black, fill=white, very thick] (9.6,1.3) rectangle (11.6,1.9);
\coordinate[align=center,text width=1,label=center:{\parbox{2cm}{\centering Enforcer}}] (R) at (10.6,1.6);

\draw [draw=black, fill=white, very thick] (9.6,0.6) rectangle (11.6,1.2);
\coordinate[align=center,text width=1,label=center:{\parbox{2cm}{\centering Dashboard}}] (R) at (10.6,0.9);

\draw [decorate,very thick,decoration={brace,amplitude=5pt,raise=4pt}]
  (6.4,2.6) -- (11.8,2.6) node[midway,yshift=2em]{Runtime Verification Containers};

\node at (2.0,-0.5) [circle,fill,inner sep=1.5pt]{};
\node at (2.2,-0.5) [circle,fill,inner sep=1.5pt]{};
\node at (2.4,-0.5) [circle,fill,inner sep=1.5pt]{};
\node at (6.9,-0.5) [circle,fill,inner sep=1.5pt]{};
\node at (8.4,-0.5) [circle,fill,inner sep=1.5pt]{};
\node at (10.4,-0.5) [circle,fill,inner sep=1.5pt]{};
\node at (10.6,-0.5) [circle,fill,inner sep=1.5pt]{};
\node at (10.8,-0.5) [circle,fill,inner sep=1.5pt]{};

\draw[very thick] (0,-0.5) -- (12,-0.5);
\coordinate[label=above:{Pub/Sub Network}] (Net) at (4.8,-0.5);

\draw[very thick] (2.0,-0.5) -- (2.0,0.6);
\draw[very thick] (2.2,-0.5) -- (2.2,0.6);
\draw[very thick] (2.4,-0.5) -- (2.4,0.6);
\draw[very thick] (6.9,-0.5) -- (6.9,0.6);
\draw[very thick] (8.4,-0.5) -- (8.4,0.6);
\draw[very thick] (10.4,-0.5) -- (10.4,0.6);
\draw[very thick] (10.6,-0.5) -- (10.6,0.6);
\draw[very thick] (10.8,-0.5) -- (10.8,0.6);

\end{tikzpicture}
\end{center}
\caption{Runtime verification containers on a publish/subscribe network}
\label{fig:intro}
\end{figure}
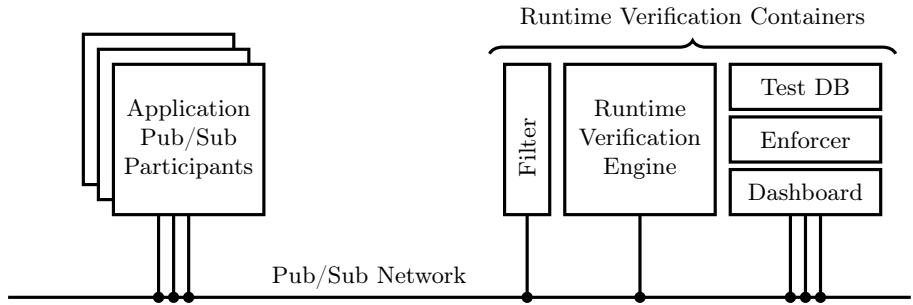

Runtime Verification (RV) is an attractive technology for the dynamic verification and monitoring of pub/sub architectures. Its inherent advantage lies in its ability to fit into the pub/sub paradigm by acting as both a subscriber and a publisher. As a subscriber, online runtime verification tools can continuously monitor message flows by subscribing to relevant topics using a proper interface. This subscriber role allows for real-time message content and behavior analysis, ensuring adherence to predefined properties and requirements. Additionally, runtime verification tools can leverage their publisher role to detect property violations and anomalies at runtime, facilitating proactive error detection and enforcement. This dual subscriber-publisher functionality enables comprehensive monitoring and analysis, safeguarding system-level reliability and correctness, especially in dynamic environments. Nevertheless, deploying and maintaining pub/sub participants is a non-trivial task extending to runtime verification participants. This is a significant challenge in front of a streamlined runtime verification framework targeting pub/sub networks.

As a solution, containerization of pub/sub network and runtime verification participants seems very attractive for overcoming deployment and maintenance problems. Containers offer simplified deployments, enhanced portability, efficient scaling, and rapid updates, ultimately contributing to a more reliable, adaptable, and efficient communication infrastructure. Therefore, we propose and study \emph{runtime verification containers} in this paper. Figure~\ref{fig:intro} illustrates RV containers encapsulating runtime verification functionality and receiving messages from subscribed topics. Runtime monitors may be designed to listen to message synchronizers and publish verdict messages to central test results databases or dashboards.  However, containerized network performance is a critical point for successfully implementing this approach. Therefore, we study runtime verification containers from a performance focus in this paper, which makes the following contributions:
\begin{itemize}
\item We perform basic performance measurements on containerized publisher/subscriber networks to establish a baseline.
\item We containerize an existing specification-based runtime verification tool and benchmark generator.
\item We compare runtime verification monitoring performance over the target pub/sub networks over several container network architectures.
\item We demonstrate the application of runtime verification containers on a real-world autonomous driving stack.
\end{itemize}

\clearpage
\section{Background}
\label{sec:preliminaries}

This section provides a general background on publish/subscribe networks, software containerization, and academic literature relevant to our study.

\subsection{Publish/Subscribe Networks}

In this paper, we mainly target DDS\footnote{\url{https://www.omg.org/dds/}} and Zenoh\footnote{\url{https://zenoh.io}} communication protocols based on the publish/subscribe paradigm, used primarily for implementing complex real-time systems and very popular in robotics.

\paragraph{Data Distribution Service (DDS)} is an open data-centric publish/subscribe protocol communications for real-time and embedded systems standardized by the Object Management Group (OMG). Unlike traditional message-passing systems, DDS boasts a fully decentralized architecture. This means network participants can discover each other automatically through a dynamic discovery service, eliminating the need for a central broker. This, combined with the interoperability of different DDS implementations, ensures seamless communication across diverse systems. However, it's important to note that the discovery service itself can become a performance bottleneck when managing a large number of participants. DDS utilizes the Common Data Representation (CDR) format for serializing messages over the wire. DDS has been used as the default middleware of the Robot Operating System (ROS), a popular framework for developing robots in academia.

\paragraph{Eclipse Zenoh} is a recently introduced open-source communication protocol that leverages the publisher/subscriber and client-server paradigms, operating under the Eclipse Foundation. It aims to address limitations inherent in existing protocols like DDS by enabling the construction of highly configurable network topologies. These topologies can extend beyond a single physical location, allowing the construction of highly scalable distributed systems over the cloud and edge devices in a unified manner. Zenoh does not dictate a particular serialization format, and users can use any serialization format as long as publishers and subscribers agree on the format. This flexibility seems to be Zenoh's main advantage compared to more established alternatives, along with its modern library-oriented design.
 
\bigskip\noindent The pub/sub networks landscape is not limited to DDS implementations and Zenoh. Other pub/sub protocols like MQTT and Redis are popular, especially for applications that do not require high performance but flexibility.
 
\subsection{Modern Container Technology}

Containers are light virtualization solutions whose origins can be traced back to the \texttt{chroot} utility in the early UNIX systems. The support for isolation features, such as \texttt{namespaces} and \texttt{cgroups}, introduced in the Linux kernel has been a turning point in the early 2000s, and several containerization technologies have emerged since then~\cite{bentaleb2022containerization}. Docker containers~\cite{merkel2014docker}, later standardized by the Open Container Initiative (OCI) initiative, is the most popular containerization technology, providing a large ecosystem of user-friendly tools and conventions.

For CPU-intensive tasks, containers can reach a near-native performance. It is possible to restrict computing resources individually per container, easing deployment for resource-constrained environments. However, container network isolation can introduce significant overhead for inter-container communication tasks. Therefore, modern container runtimes support multiple networking modes to trade off network isolation and performance~\cite{claassen2016linux, kozhirbayev2017performance}.

\subsection{Related Work}

Different runtime verification frameworks have been proposed in the literature, providing a variety of runtime monitors and checks for publish/subscribe networks, especially for the Robot Operating System (ROS). The first version of ROS employed a custom pub/sub protocol, whereas ROS 2 uses the DDS protocol under the hood. The ROSRV framework~\cite{huang2014rosrv}, together with ~\cite{adam2014towards}, is one of the earliest attempts to apply runtime monitoring on ROS-based robotic applications. Monitoring has been mentioned as a promising direction for robotics in~\cite{abbas2018special}. The work presented in~\cite{ferrando2020rosmonitoring}. The framework in~\cite{stadler2023romosu} focuses on the supporting components for monitoring. Code generation from requirements and specifications is another approach to generate runtime monitors~\cite{perez2022monitoring}. Temporal logic monitoring on ROS has been studied in~\cite{yamaguchi2024rtamt}. These works, however, lack a viable strategy to be integrated into industrial workflows. Finally, a comprehensive survey on runtime verification in the ROS ecosystem is recently presented in~\cite{caldas2024runtime}.

Containerization and microservices architectures have gained significant traction in robotic applications~\cite{xia2018microservice,lumpp2021container,dawarka2022building}. Similarly, the automotive industry has been moving to software-defined solutions exemplified in the industrial consortiums such as Eclipse SDV\footnote{\url{https://sdv.eclipse.org/}} and SOAFEE\footnote{\url{https://www.soafee.io/}}. The overhead of system virtualization in robotic and automotive applications has been studied for another automotive application in~\cite{wen2023bare}. The granularity of containerization is also a concern in these discussions, and a comparative study is presented in~\cite{betz2024containerized}. Another important topic is the real-time and mixed-criticality containers, as standard containers do not have real-time guarantees.  In~\cite{lumpp2023enabling}, authors study real-time schedulers for Kubernetes orchestration for robotic applications. However, their container and system monitoring is limited to basic health checks, unlike functional system properties that a runtime verification tool can provide. 

\clearpage
\section{Containerization}

This section presents our containerization approach to an existing specification-based runtime verification tool. Reelay is an online runtime verification library to monitor message streams against past-time metric and first-order temporal logic specifications~\cite{ulus2019online}. While existing works primarily focused on demonstrating the performance of online monitoring over logfile-based Timescales benchmarks~\cite{ulus2019timescales}, we are interested in several containerized architectures that integrate runtime verification functionality and their performance, as explained in the following. 

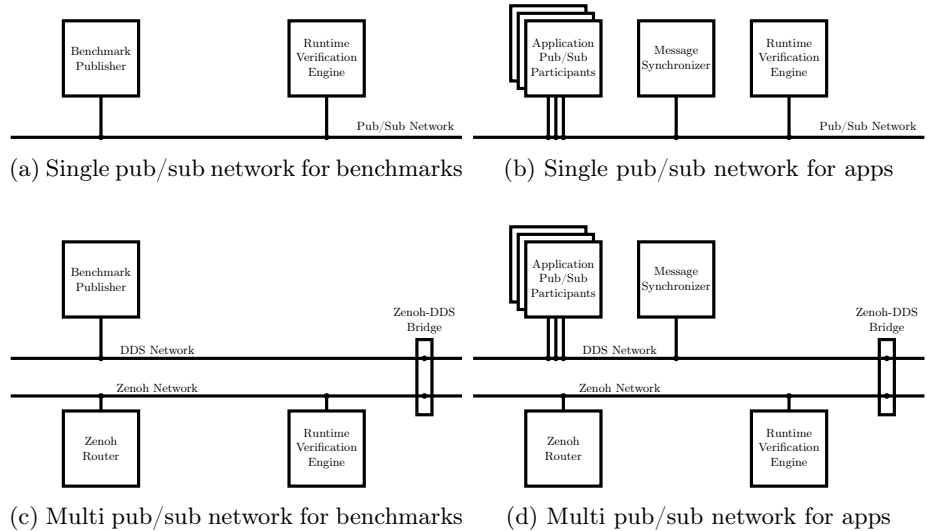
\begin{figure}[!b]
\centering
\begin{subfigure}[t]{0.495\textwidth}
\centering
\begin{tikzpicture}[scale=0.5, transform shape]

\draw [draw=black, fill=white, very thick] (1.4,0.6) rectangle (3.4,2.6);
\coordinate[align=center,text width=1,label=center:{\parbox{2cm}{\centering Benchmark\\Publisher}}] (R) at (2.4,1.6);

\draw [draw=black, fill=white, very thick] (7.4,0.6) rectangle (9.4,2.6);
\coordinate[align=center,text width=1,label=center:{\parbox{2cm}{\centering Runtime\\Verification\\Engine}}] (R) at (8.4,1.6);

\node at (2.4,-0.5) [circle,fill,inner sep=1.5pt]{};
\node at (8.4,-0.5) [circle,fill,inner sep=1.5pt]{};

\draw[very thick] (2.4,-0.5) -- (2.4,0.6);
\draw[very thick] (8.4,-0.5) -- (8.4,0.6);

\draw[very thick] (0,-0.5) -- (12,-0.5);
\coordinate[label=above:{Pub/Sub Network}] (Net) at (10.5,-0.5);
\end{tikzpicture}
\caption{Single pub/sub network for benchmarks}
\end{subfigure}
\begin{subfigure}[t]{0.495\textwidth}
\centering
\begin{tikzpicture}[scale=0.5, transform shape]
\draw [draw=black, fill=white, very thick] (1,1) rectangle (3,3);
\draw [draw=black, fill=white, very thick] (1.2,0.8) rectangle (3.2,2.8);
\draw [draw=black, fill=white, very thick] (1.4,0.6) rectangle (3.4,2.6);
\coordinate[align=center,text width=1,label=center:{\parbox{2cm}{\centering Application\\Pub/Sub\\Participants}}] (R) at (2.4,1.6);

\draw [draw=black, fill=white, very thick] (4.4,0.6) rectangle (6.4,2.6);
\coordinate[align=center,text width=1,label=center:{\parbox{2cm}{\centering Message\\Synchronizer}}] (R) at (5.4,1.6);

\draw [draw=black, fill=white, very thick] (7.4,0.6) rectangle (9.4,2.6);
\coordinate[align=center,text width=1,label=center:{\parbox{2cm}{\centering Runtime\\Verification\\Engine}}] (R) at (8.4,1.6);

\node at (2.0,-0.5) [circle,fill,inner sep=1.5pt]{};
\node at (2.2,-0.5) [circle,fill,inner sep=1.5pt]{};
\node at (2.4,-0.5) [circle,fill,inner sep=1.5pt]{};
\node at (5.4,-0.5) [circle,fill,inner sep=1.5pt]{};
\node at (8.4,-0.5) [circle,fill,inner sep=1.5pt]{};

\draw[very thick] (2.0,-0.5) -- (2.0,0.6);
\draw[very thick] (2.2,-0.5) -- (2.2,0.6);
\draw[very thick] (2.4,-0.5) -- (2.4,0.6);
\draw[very thick] (5.4,-0.5) -- (5.4,0.6);
\draw[very thick] (8.4,-0.5) -- (8.4,0.6);

\draw[very thick] (0,-0.5) -- (12,-0.5);
\coordinate[label=above:{Pub/Sub Network}] (Net) at (10.5,-0.5);
\end{tikzpicture}
\caption{Single pub/sub network for apps}
\end{subfigure}

\vspace{0.5cm}

\begin{subfigure}[t]{0.495\textwidth}
\centering
\begin{tikzpicture}[scale=0.5, transform shape]
\draw [draw=black, fill=white, very thick] (1.4,0.6) rectangle (3.4,2.6);
\coordinate[align=center,text width=1,label=center:{\parbox{2cm}{\centering Benchmark\\Publisher}}] (R) at (2.4,1.6);

\draw [draw=black, fill=white, very thick] (7.4,-1.9) rectangle (9.4,-3.9);
\coordinate[align=center,text width=1,label=center:{\parbox{2cm}{\centering Runtime\\Verification\\Engine}}] (R) at (8.4,-2.9);

\draw [draw=black, fill=white, very thick] (1.4,-1.9) rectangle (3.4,-3.9);
\coordinate[align=center,text width=1,label=center:{\parbox{2cm}{\centering Zenoh\\Router}}] (Router) at (2.4,-2.9);

\draw [draw=black, fill=white, very thick] (10.8,0) rectangle (11.2,-2);
\coordinate[align=center,text width=1,label=center:{\parbox{2cm}{\centering Zenoh-DDS Bridge}}] (Router) at (11,0.5);

\node at (2.4,-0.5) [circle,fill,inner sep=1.5pt]{};
\node at (8.4,-1.5) [circle,fill,inner sep=1.5pt]{};
\node at (11,-0.5) [circle,fill,inner sep=1.5pt]{};
\node at (11,-1.5) [circle,fill,inner sep=1.5pt]{};
\node at (2.4,-1.5) [circle,fill,inner sep=1.5pt]{};

\draw[very thick] (2.4,-0.5) -- (2.4,0.6);
\draw[very thick] (8.4,-1.5) -- (8.4,-1.9);
\draw[very thick] (2.4,-1.5) -- (2.4,-1.9);

\draw[very thick] (0,-0.5) -- (12,-0.5);
\coordinate[label=above:{DDS Network}] (NetDDS) at (3.9,-0.5);

\draw[very thick] (0,-1.5) -- (12,-1.5);
\coordinate[label=above:{Zenoh Network}] (NetZenoh) at (3.9,-1.5);
\end{tikzpicture}
\caption{Multi pub/sub network for benchmarks}
\end{subfigure}
\begin{subfigure}[t]{0.495\textwidth}
\begin{tikzpicture}[scale=0.5, transform shape]
\draw [draw=black, fill=white, very thick] (1,1) rectangle (3,3);
\draw [draw=black, fill=white, very thick] (1.2,0.8) rectangle (3.2,2.8);
\draw [draw=black, fill=white, very thick] (1.4,0.6) rectangle (3.4,2.6);
\coordinate[align=center,text width=1,label=center:{\parbox{2cm}{\centering Application\\Pub/Sub\\Participants}}] (R) at (2.4,1.6);

\draw [draw=black, fill=white, very thick] (4.4,0.6) rectangle (6.4,2.6);
\coordinate[align=center,text width=1,label=center:{\parbox{2cm}{\centering Message\\Synchronizer}}] (R) at (5.4,1.6);

\draw [draw=black, fill=white, very thick] (7.4,-1.9) rectangle (9.4,-3.9);
\coordinate[align=center,text width=1,label=center:{\parbox{2cm}{\centering Runtime\\Verification\\Engine}}] (R) at (8.4,-2.9);

\draw [draw=black, fill=white, very thick] (1.4,-1.9) rectangle (3.4,-3.9);
\coordinate[align=center,text width=1,label=center:{\parbox{2cm}{\centering Zenoh\\Router}}] (Router) at (2.4,-2.9);

\draw [draw=black, fill=white, very thick] (10.8,0) rectangle (11.2,-2);
\coordinate[align=center,text width=1,label=center:{\parbox{2cm}{\centering Zenoh-DDS Bridge}}] (Router) at (11,0.5);

\node at (2.0,-0.5) [circle,fill,inner sep=1.5pt]{};
\node at (2.2,-0.5) [circle,fill,inner sep=1.5pt]{};
\node at (2.4,-0.5) [circle,fill,inner sep=1.5pt]{};
\node at (5.4,-0.5) [circle,fill,inner sep=1.5pt]{};
\node at (8.4,-1.5) [circle,fill,inner sep=1.5pt]{};
\node at (11,-0.5) [circle,fill,inner sep=1.5pt]{};
\node at (11,-1.5) [circle,fill,inner sep=1.5pt]{};
\node at (2.4,-1.5) [circle,fill,inner sep=1.5pt]{};

\draw[very thick] (2.0,-0.5) -- (2.0,0.6);
\draw[very thick] (2.2,-0.5) -- (2.2,0.6);
\draw[very thick] (2.4,-0.5) -- (2.4,0.6);
\draw[very thick] (5.4,-0.5) -- (5.4,0.6);
\draw[very thick] (8.4,-1.5) -- (8.4,-1.9);
\draw[very thick] (2.4,-1.5) -- (2.4,-1.9);

\draw[very thick] (0,-0.5) -- (12,-0.5);
\coordinate[label=above:{DDS Network}] (NetDDS) at (3.9,-0.5);

\draw[very thick] (0,-1.5) -- (12,-1.5);
\coordinate[label=above:{Zenoh Network}] (NetZenoh) at (3.9,-1.5);
\end{tikzpicture}
\caption{Multi pub/sub network for apps}
\end{subfigure}
\caption{Container network architectures considered in the paper}
\label{fig:architecture}
\end{figure}

\subsection{Target Architectures}

The core component in our architectures is the runtime verification engine container. This container encapsulates an executable that utilizes the Reelay library and the relevant pub/sub API for the chosen communication protocol, such as DDS or Zenoh. The executable subscribes to the specified topics and processes each input message using the RV engine before publishing the current verdict message. Reelay monitors read JSON strings by default, and both DDS and Zenoh networks allow strings to be sent and received. Figure~\ref{fig:architecture} illustrates the container architectures explored in this paper for performance evaluation and case studies. We start with a simple benchmarking setup in Figure~\ref{fig:architecture}(a), where the benchmark publisher component publishes simulated data messages in JSON format on the specific topic.

The second setup in Figure~\ref{fig:architecture}(b) illustrates a typical application setup that involves multiple participants, and the RV container is attached to the network as another participant. Publish/subscribe networks usually employ asynchronous communication, whereas runtime verification tools such as Reelay demand a strictly synchronized stream of messages. Therefore, we need an adapter component, the message synchronizer, between the RV container and the publish/subscribe network. This component combines and synchronizes messages from the subscribed topics into a single topic for the RV container to subscribe to. Our case study in Section~\ref{sec:case-study} details this approach's implementation and current limitations.

Figure~\ref{fig:architecture}(c) and Figure~\ref{fig:architecture}(d) illustrates our multi-network setup for benchmarking and applications. Given the widespread adoption of the DDS protocol and the new promise of Zenoh, we mainly focus on the DDS-to-Zenoh case where the system resides on DDS and the runtime verification component on Zenoh. In these setups, we use a bridge component to translate messages from one to another. Although this adds some overhead, we consider this a common scenario when working with containerized pub/sub networks. Finally, on native Zenoh networks, participants can discover each other using multicast. This feature is unavailable for containers; therefore, a Zenoh router is necessary unless participants know each other by network addresses.

\subsection{Container Networking Modes}

Networking overhead among containers can be prohibitively large for high frequency, high throughput applications. Therefore, container runtimes support several networking modes to enable users to trade off according to their performance and isolation needs. Docker containers implement several networking modes: \texttt{bridge}, \texttt{host}, \texttt{ipvlan}, \texttt{macvlan}, and direct one-to-one connection between containers.

\subsection{Container Orchestration}

We use the Docker Compose (\texttt{docker-compose}) utility to orchestrate multi-container applications. Docker Compose tool simplifies managing applications that rely on multiple containers working together.  It enables the declarative specification of services, network configurations, data volumes, and other dependencies within a singular YAML file, conventionally named \texttt{compose.yml}. This is especially useful when we have multiple container setup configuration as we study in this paper.

\subsection{Multi-stage Container Build}

Multi-stage container build is a powerful technique for keeping container images small and efficient. Since containers are often downloaded online, container size is crucial in optimizing deployment times and minimizing bandwidth usage. This approach achieves this by separating the build and runtime environments within a single container file. Developers can leverage this separation to include only essential dependencies and artifacts in the final image, eliminating unnecessary build tools and intermediate files. Therefore, multi-stage builds can significantly reduce the overall image size by discarding these elements after the build phase. This translates to more lightweight and efficient containers for distribution, leading to faster deployments and lower bandwidth consumption.

\section{Containerized Network Performance Evaluation}
\label{sec:evaluation}

Throughput and latency measurements are two primary metrics used to demonstrate the performance of networked applications. Network throughput is the rate at which data successfully travels across a network connection, typically measured in messages per second or megabits per second. Network latency is often defined as half of the total duration for a message round-trip --- that is, publishing a message for another participant, and the participant publishes another message immediately upon receiving the first message. These measurements demonstrate the network's actual data-carrying capacity and responsiveness limits. Throughput and latency depend on various factors impacting real-world data transfer, such as network congestion, protocol overhead, and processing delays.

The arrival guarantee for pub/sub messages might be an important concern for safety-critical runtime verification applications where data loss could have catastrophic consequences. Therefore, we use DDS with \texttt{Reliable+KeepAll} quality of service (QoS) settings that guarantee that all published messages are delivered to subscribing participants. Zenoh similarly provides similar reliable QoS settings, and we activate those guarantees in our experiments. 
Protocols' versions and settings are described as follows. For the DDS experiments, we use Eclipse CycloneDDS implementation (\texttt{cyclonedds v0.10}) from distribution repos. For the Zenoh experiments, we use the C library bindings (\texttt{zenoh-c v0.10.2-rc}) and the Zenoh-DDS bridge plugin of the same Zenoh version. All benchmarks are conducted on a computer with the following specifications:

\vspace{0.2cm}
\begin{itemize}
  \item \textbf{CPU:} Intel(R) Core(TM) i7-10750H CPU @ 2.60GHz, 12 cores
  \item \textbf{Memory:} 32GB RAM
  \item \textbf{GPU:} NVIDIA GeForce RTX 2060
  \item \textbf{Network:} Intel Corporation Wi-Fi 6 AX201 \& Realtek RTL8111/8168/8411 PCI Express Gigabit Ethernet Controller
  \item \textbf{Kernel:} Linux (5.15.0-105-generic)
  \item \textbf{Operating System:} Ubuntu Linux 22.04
\end{itemize}
\vspace{0.2cm}

We present our baseline performance experiments for Zenoh and DDS protocols given in Section~\ref{sec:baseperf}. Our experiments for runtime verification containers in Section~\ref{sec:rvperf} demonstrate performance compared to the baseline and not-networked settings.

\subsection{Baseline Experiments in Containers}
\label{sec:baseperf}

Our baseline DDS experiments use the industry-standard \texttt{ddsperf} utility for  DDS performance measurements. For baseline Zenoh cases, we employ custom Zenoh applications from the Zenoh C repository\footnote{\url{https://github.com/eclipse-zenoh/zenoh-c}} to measure both throughput and latency. These applications enable fine-grained control over message payload size and frequency, allowing comprehensive benchmarking. We containerize these applications using Docker containers and orchestrate them with the \texttt{docker-compose} utility to test with the aforementioned message parameters and container network modes. In the following, we present performance results from these experiments. 

\begin{figure}[tbh!]
\centering
\begin{subfigure}[c]{0.495\textwidth}
\centering
\caption{CycloneDDS Throughput}
\includegraphics[width=\textwidth]{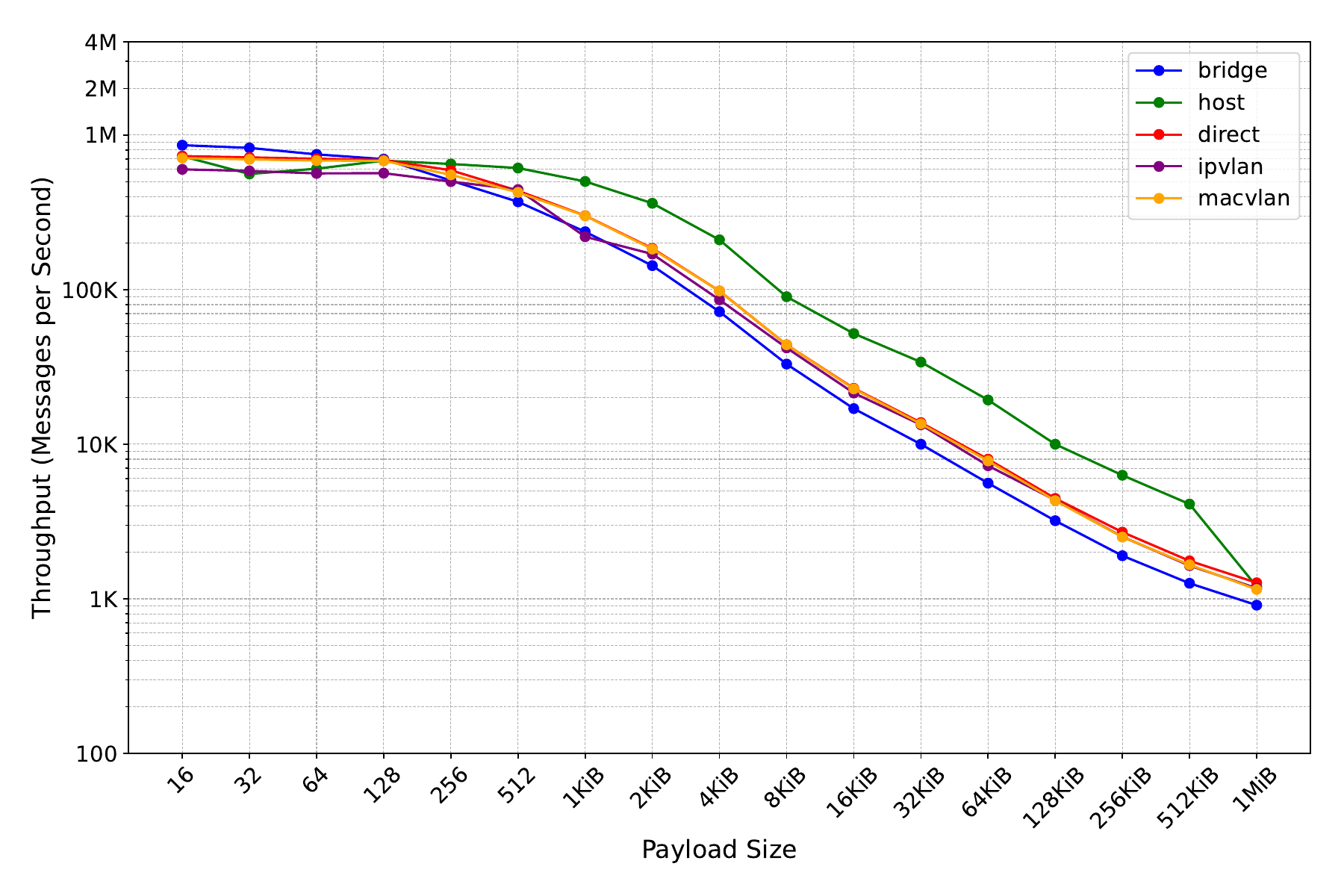}
\end{subfigure}
\begin{subfigure}[c]{0.495\textwidth}
\centering
\caption{CycloneDDS Latency}
\includegraphics[width=\textwidth]{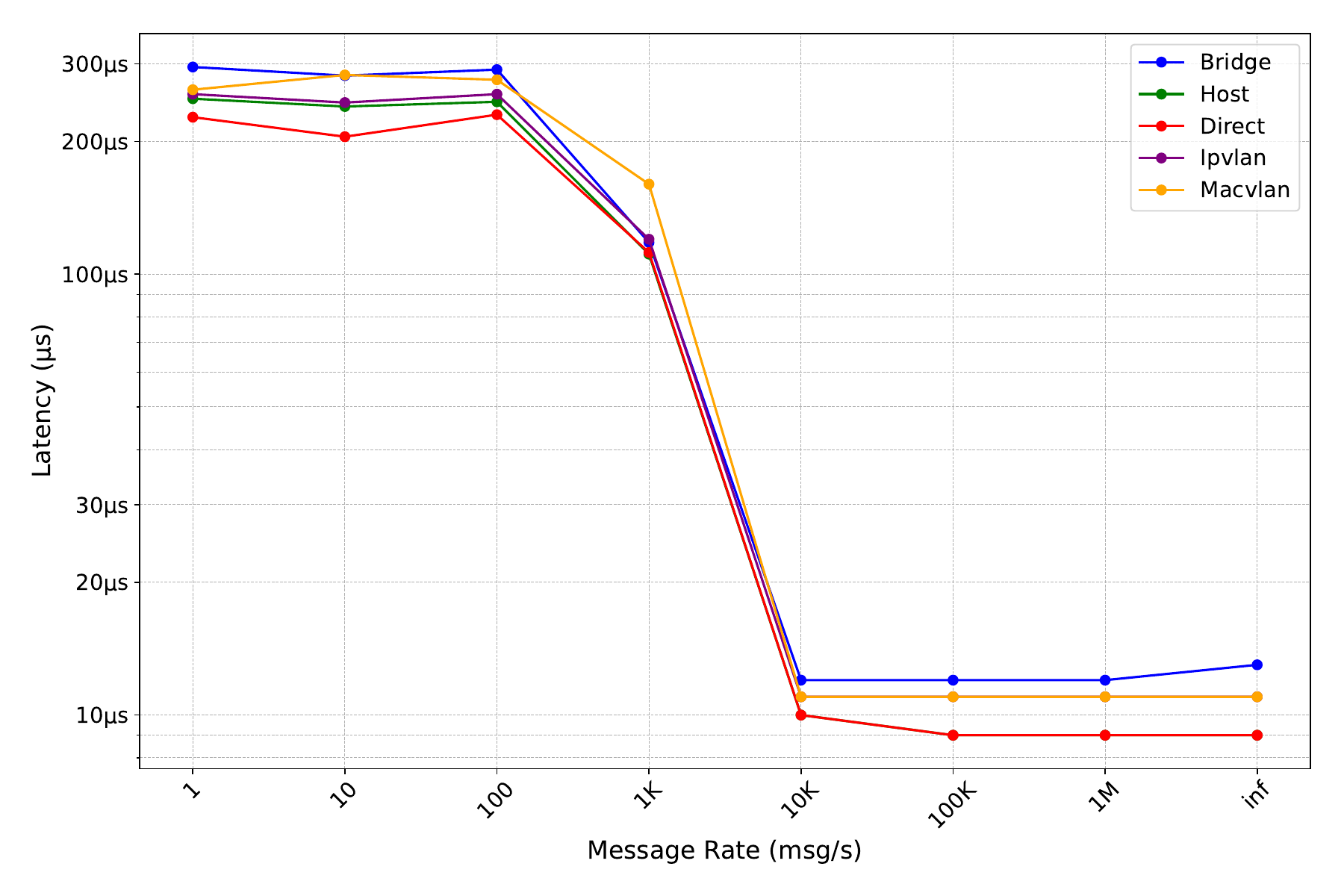}
\end{subfigure}
\begin{subfigure}[c]{0.495\textwidth}
\centering
\caption{Zenoh Throughput}
\includegraphics[width=\textwidth]{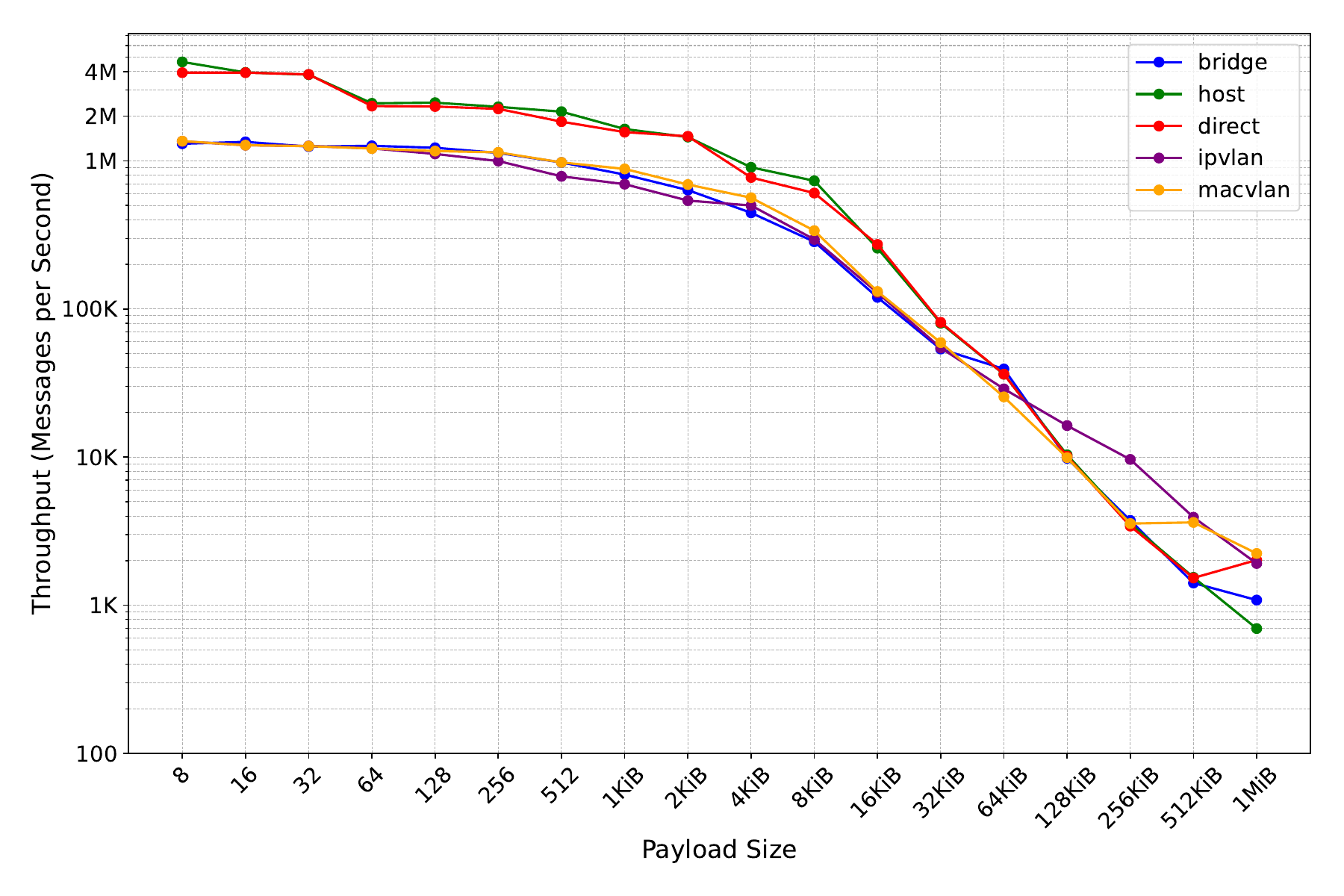}
\end{subfigure}
\begin{subfigure}[c]{0.495\textwidth}
\centering
\caption{Zenoh Latency}
\includegraphics[width=\textwidth]{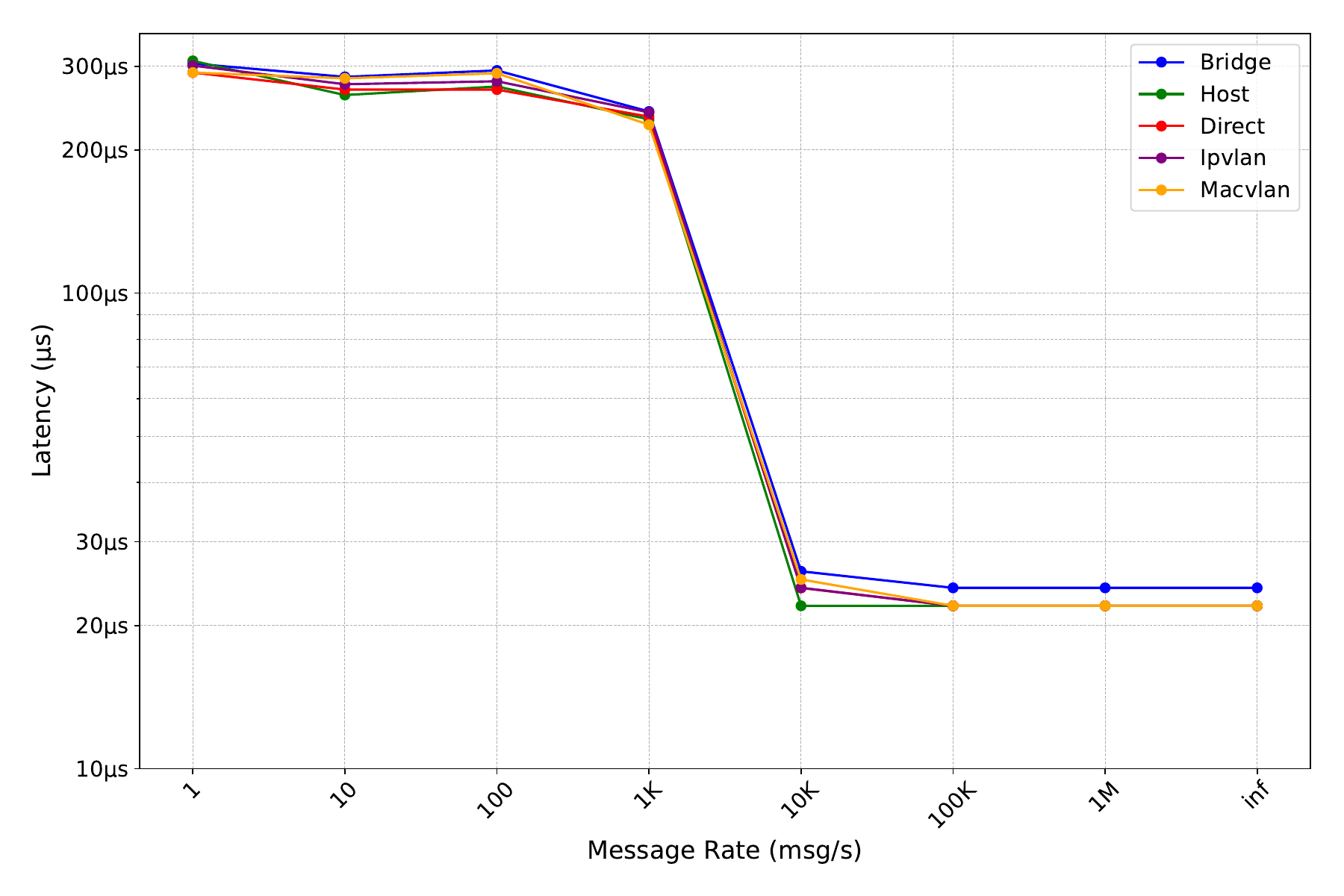}
\end{subfigure}
\begin{subfigure}[c]{0.495\textwidth}
\centering
\caption{DDS-to-Zenoh Throughput}
\includegraphics[width=\textwidth]{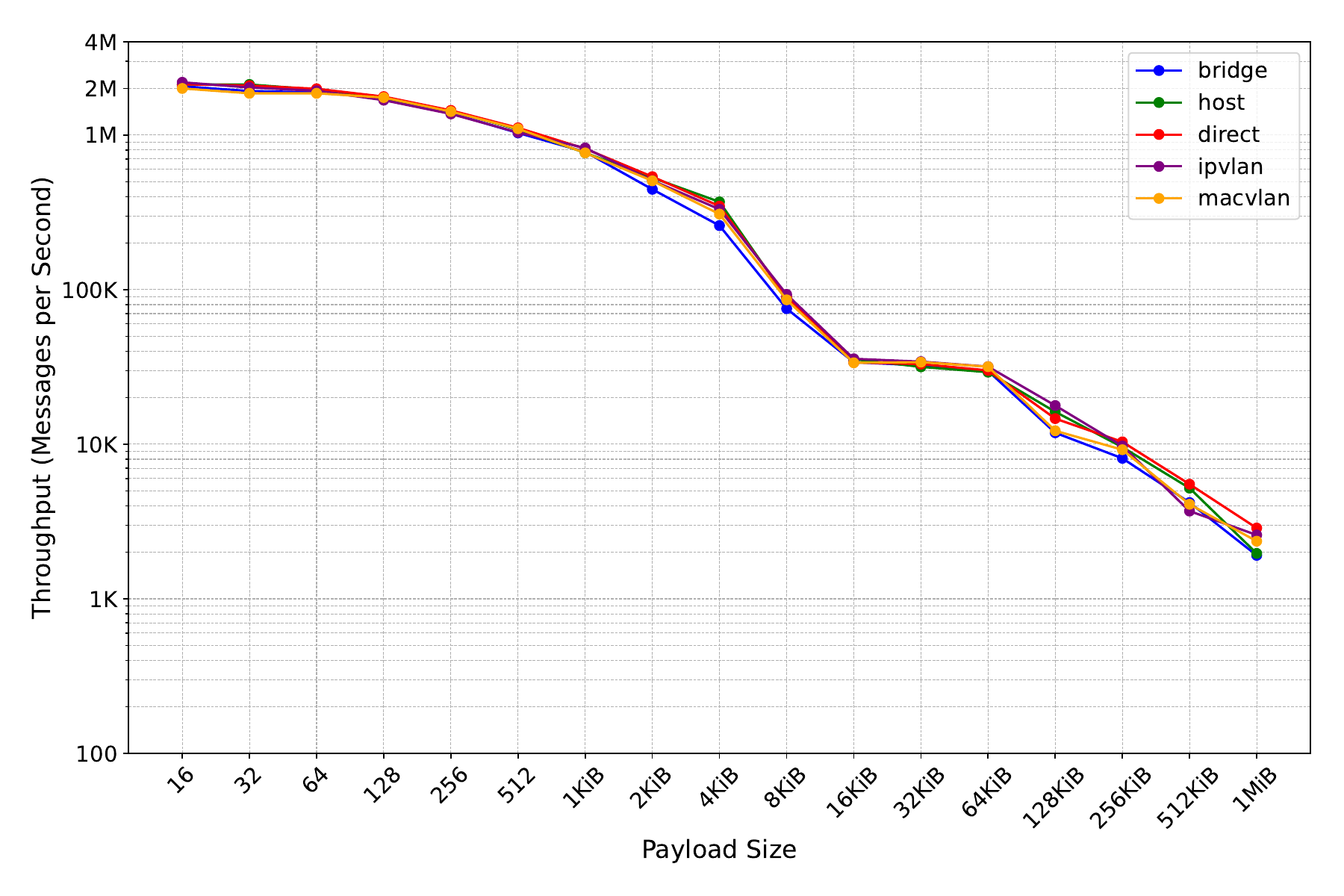}
\end{subfigure}
\begin{subfigure}[c]{0.495\textwidth}
\centering
\caption{DDS-to-Zenoh Latency}
\includegraphics[width=\textwidth]{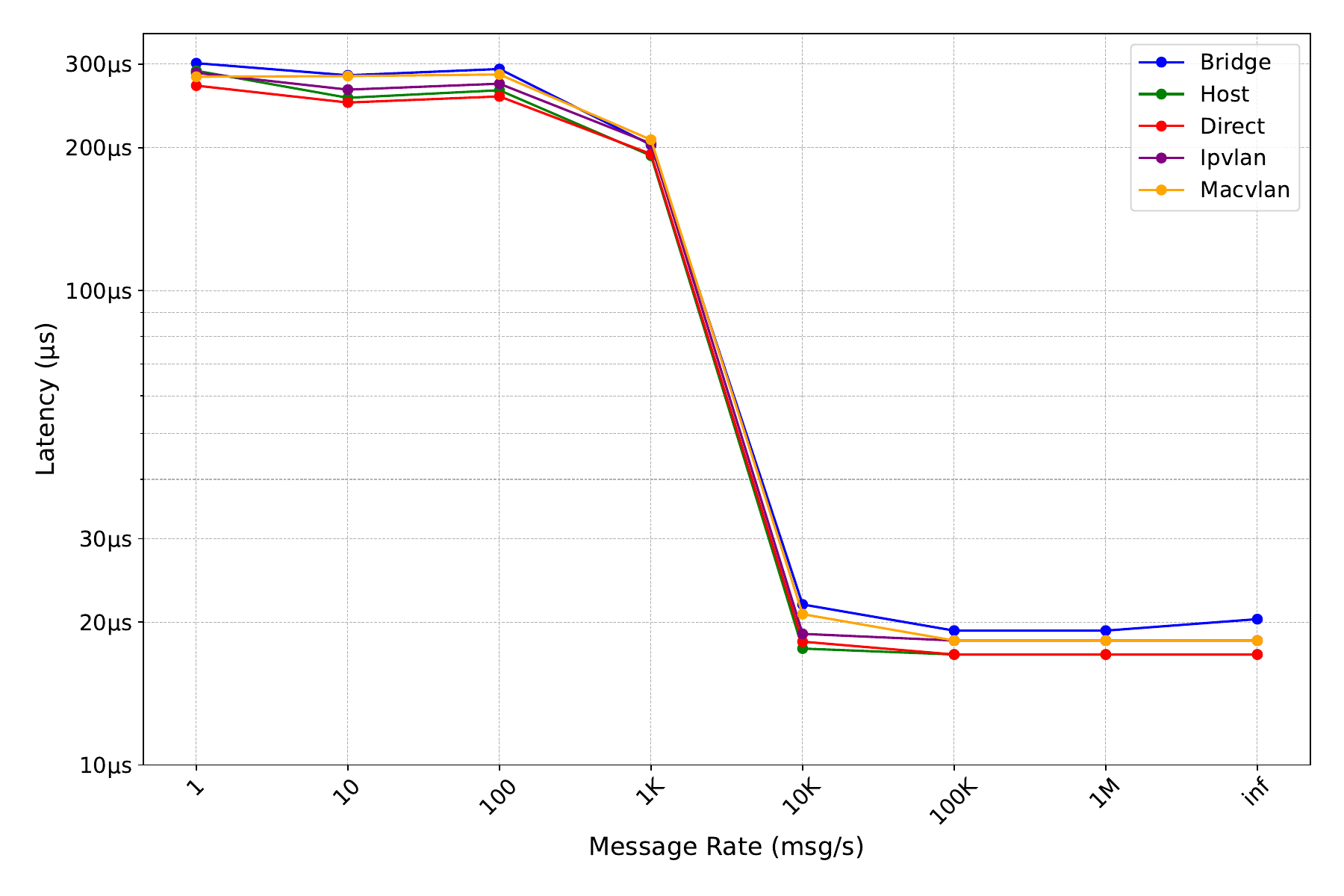}
\end{subfigure}
\caption{Baseline network performance for DDS and Zenoh pub/sub protocols}
\label{fig:baseline}
\end{figure}

Throughput measurement plots in Figure~\ref{fig:baseline}(a) and Figure~\ref{fig:baseline}(c) demonstrate the expected decrease in messages per second as the payload size increases. For very small messages, however, we observe a constant performance as the overhead of the protocol is more dominant. More curiously, container network modes considerably affect the throughput performance of containerized pub/sub networks in these benchmarks. No network isolation cases (\texttt{host}, \texttt{direct}) significantly more performant than full (\texttt{bridge}) or partial (\texttt{ipvlan}, \texttt{macvlan}) network isolation. Zenoh protocol performs better, reaching 4M msg/s rates for small messages, and sustaining 2M msg/s for medium-sized messages. However, CycloneDDS could reach 1M msg/s rates in our experiments.

Latency measurement plots in Figure~\ref{fig:baseline}(b) and Figure~\ref{fig:baseline}(d) confirm an interesting phenomenon about pub/sub networks. In these plots, we see two distinct behavior regimes depending on the frequency of messages --- low-frequency and high-frequency regimes. This phenomenon has been reported in~\cite{liang2023performance}, and our experiment reproduces it for the containerized case across all networking modes. The latency value with low-frequency messages can be as high as 300 microseconds for DDS and Zenoh, whereas this value can go down 20 microseconds when the frequency increases. The threshold seems to be between 1K and 10K messages per second in our experiments. The root cause for this sharp latency behavior change is the context switching under the operating system. The high frequency of messaging keeps the process active and does not let the operating system put it to sleep; therefore, high-frequency messaging does not need the costly operation of waking up for a new message. Also, we note that latency values for the Zenoh protocol were measured more closely among network modes. Although DDS performs better than Zenoh in terms of latency, this gap is reduced under network isolation. 

Finally, we present our benchmark results for the cross-protocol DDS-to-Zenoh case in Figure~\ref{fig:baseline}(e) and Figure~\ref{fig:baseline}(f) where subscribers from Zenoh networks can receive messages from participants from DDS networks. This case is possible with an additional participant to bridge both networks, called the Zenoh-DDS bridge. The bridge participant performs the necessary translation between networks by subscribing specified topics from one network and publish them to the other --- not to be confused with a similar yet different concept of bridge container network mode as discussed earlier. We find this case interesting as many legacy systems already use DDS-based architectures, and their migration to a new pub/sub framework would not happen immediately, even if Zenoh is a promising alternative for new applications. This is also true for us, and we acknowledge that we need to monitor DDS networks even though we prefer developing our runtime verification application over Zenoh.

\subsection{Runtime Verification Experiments in Containers}
\label{sec:rvperf}

This section describes our performance evaluation of containerized runtime verification applications. To this end, we employ the Reelay library, developed for online monitoring of Metric Temporal Logic (MTL) formulas. In our experiments, Reelay is containerized as a subscriber to listen to a single pub/sub topic. This topic delivers propositional values for atomic propositions inside formulas in a streaming fashion. On the publisher side, we employ a custom participant to publish messages one by one to the pub/sub network after reading them from Timescales benchmarks, originally used to demonstrate the performance of MTL monitoring tools. By default, Reelay expects propositional values in JSON-formatted messages, therefore we use JSON strings as message payloads. Timescales messages are small messages containing at most four propositional values, amounting to JSON strings no larger than 32 bytes in size.

\begin{figure}[b!]
\centering
\begin{subfigure}[c]{0.495\textwidth}
\centering
\caption{Reelay-DDS Latency Performance}
\includegraphics[width=\textwidth]{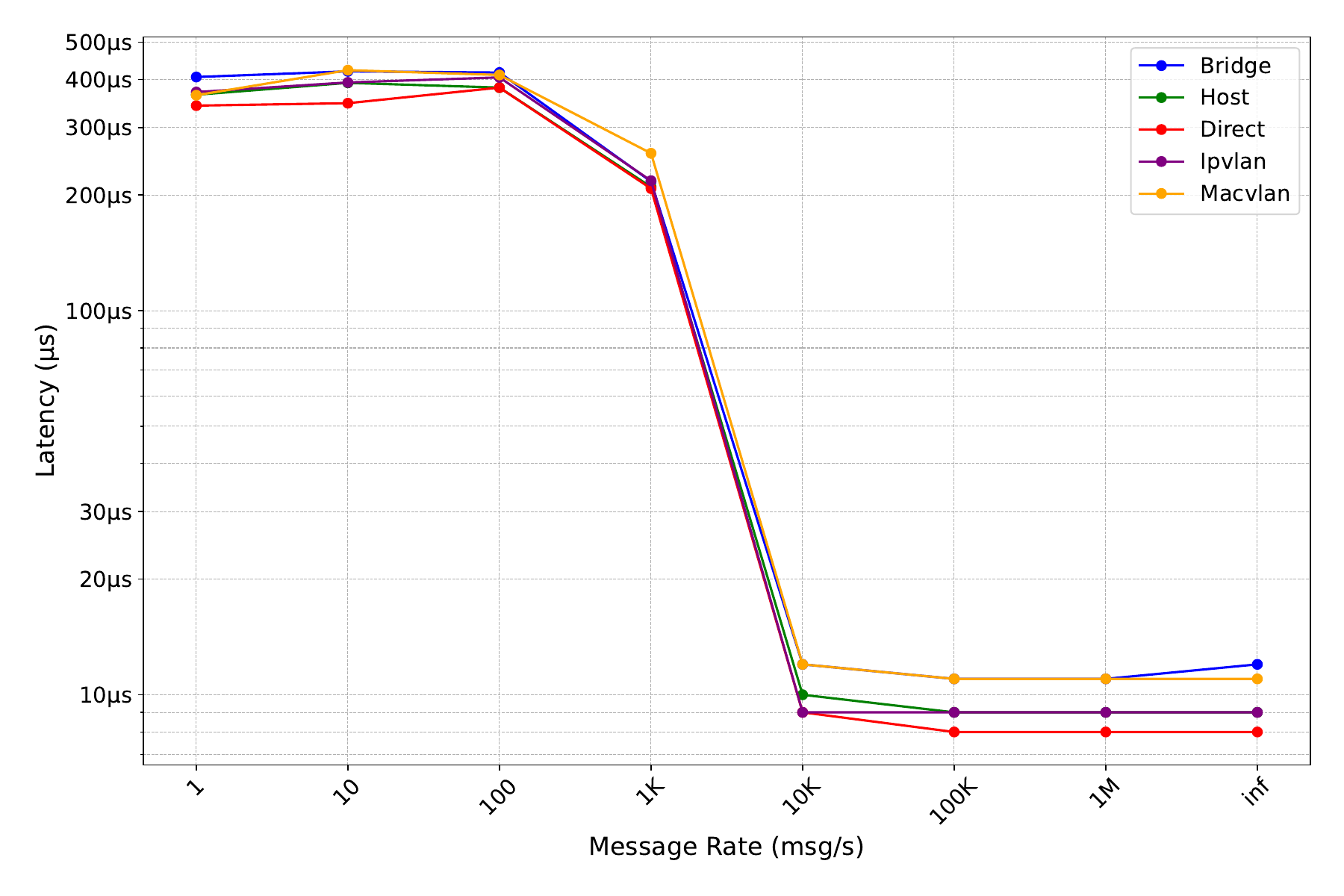}
\end{subfigure}
\begin{subfigure}[c]{0.495\textwidth}
\centering
\caption{Reelay-Zenoh Latency Performance}
\includegraphics[width=\textwidth]{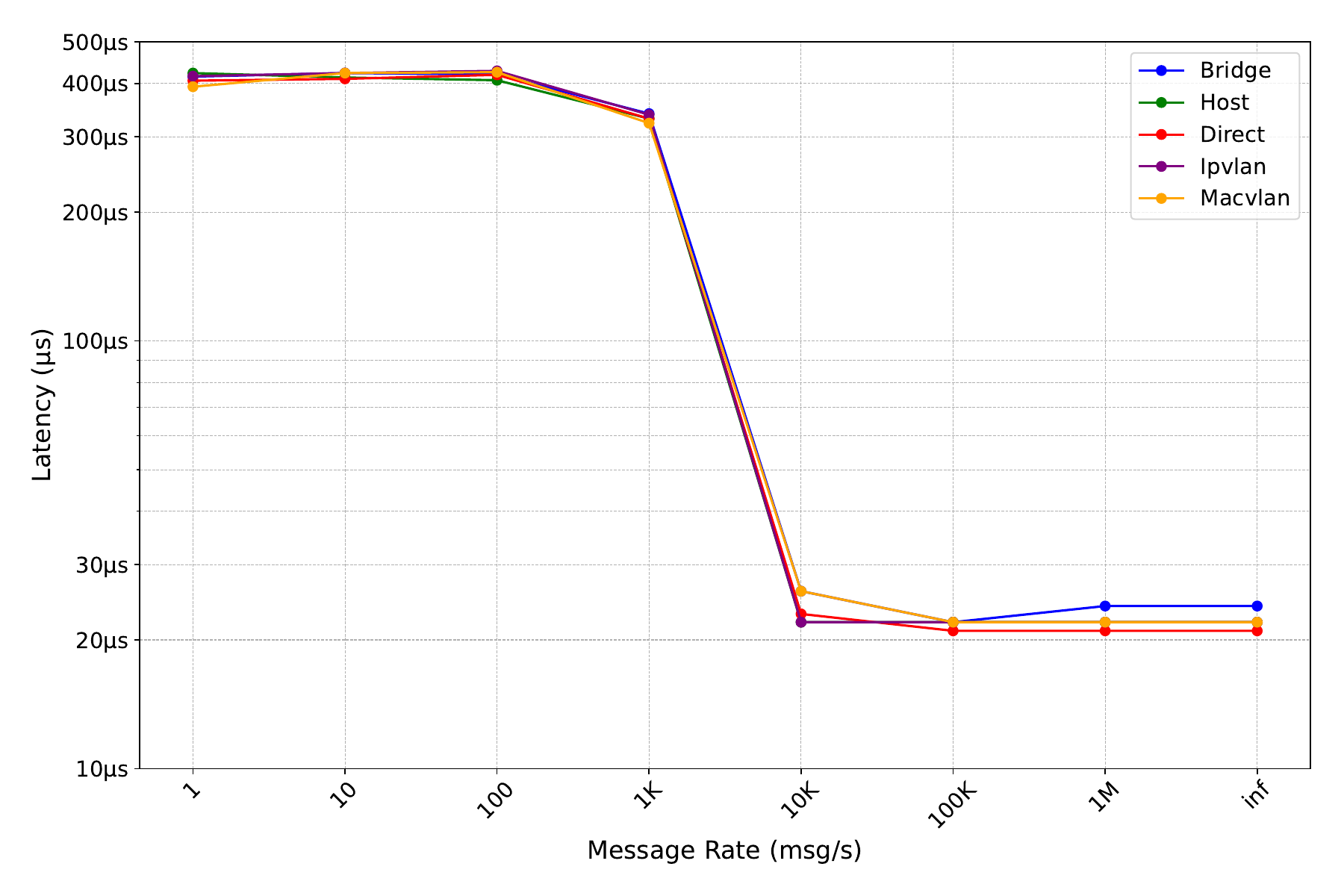}
\end{subfigure}
\caption{Latency performance for containerized runtime verification}
\label{fig:reelay_latency}
\end{figure}

\begin{table}[t!]
\centering
\caption{Runtime verification container performance on pub/sub networks}
\resizebox{\textwidth}{!}{
\begin{tabularx}{\textwidth}{>{\ttfamily\footnotesize}l*{8}{>{\raggedleft\arraybackslash\footnotesize}X}}
\toprule
& &\multicolumn{2}{r}{CycloneDDS} & & \multicolumn{2}{r}{Zenoh} & \multicolumn{2}{r}{Filesystem}\\
\cmidrule(lr){3-4} \cmidrule(lr){6-7} \cmidrule(lr){8-9}
& & Bridge & Host & & Bridge & Host & & Local\\
\midrule
AbsentAQ10    & & 1.56 & 1.38 & & 0.71 & 0.72 & & 0.39\\
AbsentAQ100   & & 1.55 & 1.35 & & 0.64 & 0.67 & & 0.36\\
AbsentAQ1000  & & 1.54 & 1.37 & & 0.65 & 0.68 & & 0.36\\
\midrule
AbsentBQR10   & & 1.54 & 1.48 & & 0.82 & 0.80 & & 0.52\\
AbsentBQR100  & & 1.49 & 1.46 & & 0.73 & 0.74 & & 0.48\\
AbsentBQR1000 & & 1.50 & 1.50 & & 0.74 & 0.74 & & 0.48\\
\midrule
AbsentBR10    & & 1.41 & 1.39 & & 0.66 & 0.70 & & 0.42\\
AbsentBR100   & & 1.39 & 1.37 & & 0.67 & 0.73 & & 0.42\\
AbsentBR1000  & & 1.40 & 1.38 & & 0.66 & 0.67 & & 0.42\\
\midrule
RecurGLB10    & & 1.35 & 1.35 & & 0.63 & 0.69 & & 0.30\\
RecurGLB100   & & 1.32 & 1.31 & & 0.58 & 0.61 & & 0.23\\
RecurGLB1000  & & 1.34 & 1.33 & & 0.56 & 0.56 & & 0.22\\
\midrule
RespondGLB10  & & 1.54 & 1.50 & & 0.83 & 0.85 & & 0.56\\
RespondGLB100 & & 1.50 & 1.47 & & 0.89 & 0.91 & & 0.46\\
RespondGLB1000& & 1.51 & 1.44 & & 0.80 & 0.89 & & 0.45\\
\midrule
AlwaysAQ10    & & 1.42 & 1.41 & & 0.66 & 0.65 & & 0.38\\
AlwaysAQ100   & & 1.41 & 1.43 & & 0.65 & 0.63 & & 0.35\\
AlwaysAQ1000  & & 1.43 & 1.39 & & 0.64 & 0.64 & & 0.34\\
\midrule
AlwaysBR10    & & 1.42 & 1.41 & & 0.67 & 0.69 & & 0.40\\
AlwaysBR100   & & 1.43 & 1.40 & & 0.74 & 0.74 & & 0.42\\
AlwaysBR1000  & & 1.44 & 1.39 & & 0.66 & 0.67 & & 0.42\\
\midrule
AlwaysBQR10   & & 1.58 & 1.50 & & 0.74 & 0.74 & & 0.51\\
AlwaysBQR100  & & 1.52 & 1.47 & & 0.77 & 0.78 & & 0.47\\
AlwaysBQR1000 & & 1.58 & 1.46 & & 0.75 & 0.74 & & 0.46\\
\midrule
RecurBQR10    & & 1.59 & 1.57 & & 0.79 & 0.77 & & 0.62\\
RecurBQR100   & & 1.58 & 1.57 & & 0.79 & 0.79 & & 0.56\\
RecurBQR1000  & & 1.57 & 1.54 & & 0.78 & 0.78 & & 0.54\\
\midrule
RespondBQR10  & & 1.79 & 1.74 & & 0.93 & 0.91 & & 0.84\\
RespondBQR100 & & 1.83 & 1.66 & & 0.88 & 0.93 & & 0.74\\
RespondBQR1000& & 1.79 & 1.71 & & 0.87 & 0.90 & & 0.74\\
\bottomrule
\end{tabularx}
}
\label{tab:reelay_throughput}
\end{table}

We containerize Reelay as standalone CycloneDDS and Zenoh participants, which subscribe to a dedicated topic name and expect JSON-formatted strings as inputs. The container requires the MTL formula to run, where the names of proportional atoms in the formula must match the name of Boolean fields in the JSON document. Upon the start, the Reelay library constructs a runtime monitor inside the container and updates the monitor state for each incoming message from the network. By design, Reelay monitors demand a strict time ordering of input messages; therefore, we need to ensure the synchronization and ordering if those propositions come from different pub/sub topics before feeding the runtime monitor.

We include our latency measurements for the runtime verification case in Figure~\ref{fig:reelay_latency}, measured as the publishing and the runtime monitor publish the current verdict back to the network. We observe an additional delay of 100 microseconds for the wake-up. It's important to note that the actual processing delay attributable to the runtime verification functionality by Reelay is likely less than 1 microsecond for these benchmarks. As a result, this processing delay is not visually noticeable in the plots. Still, this measurement remains valuable as it provides a deeper understanding for accurately predicting the overhead of runtime monitors, especially if they are employed in feedback loops like real-time processes.

Table~\ref{tab:reelay_throughput} presents pub/sub networking and containerization performance overhead relative to reading propositional values directly from the local log files. We benchmark Reelay containers over the Timescales benchmarks for DDS and Zenoh protocols and container network modes. Total processing times are presented in the table when replaying each Timescales log file through the pub/sub network. The final column shows Reelay's raw performance when accessing data directly from local files, serving as a baseline for comparison. As expected, the table shows a noticeable increase in processing time due to pub/sub networking and containerization compared to direct local file access. Also, the results suggest that this overhead might be relatively lower for MTL formulas with greater complexity. This hints at the possibility of considering runtime verification containers to host multiple properties for greater efficiency. 

These performance values are consistent with our baseline performance benchmarks. Regarding throughput, Zenoh performs better and closer to the non-networked tests than CycloneDDS. However, based on these application experiments, we have not seen convincing evidence for trading container network isolation, as two extreme networking modes exhibited similar performances. This distinction might not be critical for real-world applications, as we study in Section~\ref{sec:case-study}, especially for compute-intensive tasks.

\section{Autoware Case Study}
\label{sec:case-study}
This case study presents two example deployment setups to evaluate the applicability of runtime verification containerization in real-world scenarios. To this end, we deploy Autoware autonomous driving stack\footnote{\url{https://autoware.org/}} and the Carla simulator\footnote{\url{https://carla.org/}}.

\begin{figure}[!t]
\centering
\begin{tikzpicture}[scale=0.8, transform shape]

\draw[very thick] (2.0,-0.5) -- (2.0,0.6);
\draw[very thick] (2.2,-0.5) -- (2.2,0.6);
\draw[very thick] (2.4,-0.5) -- (2.4,0.6);
\draw[very thick] (5.4,-0.5) -- (5.4,0.6);
\draw[very thick] (8.4,-0.5) -- (8.4,0.6);
\draw[very thick] (11.4,-0.5) -- (11.4,2.6);

\draw [draw=black, fill=white, very thick] (1,1) rectangle (3,3);
\draw [draw=black, fill=white, very thick] (1.2,0.8) rectangle (3.2,2.8);
\draw [draw=black, fill=white, very thick] (1.4,0.6) rectangle (3.4,2.6);
\coordinate[align=center,text width=1,label=center:{\parbox{2cm}{\centering Autoware\\(ROS 2)}}] (R) at (2.4,1.6);

\draw [draw=black, fill=white, very thick] (4.4,0.6) rectangle (6.4,2.6);
\coordinate[align=center,text width=1,label=center:{\parbox{2cm}{\centering Message\\Synchronizer\\(ROS 2)}}] (R) at (5.4,1.6);

\draw [draw=black, fill=white, very thick] (7.4,0.6) rectangle (9.4,2.6);
\coordinate[align=center,text width=1,label=center:{\parbox{2cm}{\centering Reelay\\(DDS)}}] (R) at (8.4,1.6);

\draw [draw=black, fill=white, very thick] (10.4,1.8) rectangle (12.4,3);
\coordinate[align=center,text width=1,label=center:{\parbox{2cm}{\centering Carla Simulator}}] (R) at (11.4,2.4);

\draw [draw=black, fill=white, very thick] (10.4,0.6) rectangle (12.4,1.6);
\coordinate[align=center,text width=1,label=center:{\parbox{2cm}{\centering Carla ROS Bridge}}] (R) at (11.4,1.1);

\node at (2.0,-0.5) [circle,fill,inner sep=1.5pt]{};
\node at (2.2,-0.5) [circle,fill,inner sep=1.5pt]{};
\node at (2.4,-0.5) [circle,fill,inner sep=1.5pt]{};
\node at (5.4,-0.5) [circle,fill,inner sep=1.5pt]{};
\node at (8.4,-0.5) [circle,fill,inner sep=1.5pt]{};
\node at (11.4,-0.5) [circle,fill,inner sep=1.5pt]{};

\draw[very thick] (1.0,-0.5) -- (15,-0.5);
\coordinate[label=above left:{DDS Network}] (Net) at (15,-0.5);
\end{tikzpicture}
\caption{Container deployment setup using a single host machine}
\label{fig:setup1}
\end{figure}
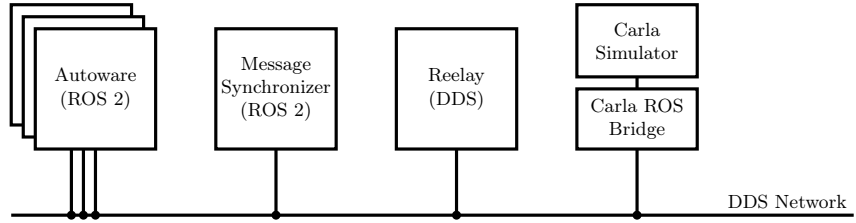

Autoware is a large open-source ROS2-based software stack for autonomous vehicles, compromising more than 250 ROS packages essential to an autonomous vehicle's sensing, perception, localization, planning, and control aspects. A typical setup of Autoware can launch a network with nearly 100 DDS participants performing compute-intensive tasks like object detection and collision avoidance. On the other hand, Carla is an autonomous driving simulator based on Unreal Engine. By default, the simulator implements its own client-server architecture and does not use any pub/sub network internally. However, the simulator may use its C++ and Python APIs to communicate with external programs. Both software stacks are resource-heavy for compiling, deploying, and executing; hence, they present certain challenges when setting up an autonomous driving testing stack. Therefore, containers and containerization technology seem very useful in managing such complexity. 

Our first deployment setup, illustrated in Figure~\ref{fig:setup1}, aims to run Autoware with the Carla simulator and Reelay runtime verification engine on the same host machine. This simple setup minimizes the need for network overhead, which can be beneficial for initial development and testing stages where performance optimization might not be the primary concern. However, it also requires a powerful workstation and may not scale well for complex scenarios with demanding resource requirements. We also note that the maximum number of participants per DDS network is limited to 120. Therefore, this setup is already at its limits.

\begin{figure}[!t]
\centering
\begin{tikzpicture}[scale=0.8, transform shape]

\draw[very thick] (5.0,-0.5) -- (5.0,0.6);
\draw[very thick] (5.2,-0.5) -- (5.2,0.6);
\draw[very thick] (5.4,-0.5) -- (5.4,0.6);
\draw[very thick] (8.4,-0.5) -- (8.4,0.6);
\draw[very thick] (8.4,-1.5) -- (8.4,-3.9);
\draw[very thick] (5.4,-1.5) -- (5.4,-1.9);
\draw[very thick] (2.4,-1.5) -- (2.4,-1.9);

\draw [draw=black, fill=white, very thick] (4,1) rectangle (6,3);
\draw [draw=black, fill=white, very thick] (4.2,0.8) rectangle (6.2,2.8);
\draw [draw=black, fill=white, very thick] (4.4,0.6) rectangle (6.4,2.6);
\coordinate[align=center,text width=1,label=center:{\parbox{2cm}{\centering Autoware\\(ROS 2)}}] (R) at (5.4,1.6);

\draw [draw=black, fill=white, very thick] (7.4,0.6) rectangle (9.4,2.6);
\coordinate[align=center,text width=1,label=center:{\parbox{2cm}{\centering Message\\Synchronizer\\(ROS 2)}}] (R) at (8.4,1.6);

\draw [draw=black, fill=white, very thick] (4.4,-1.9) rectangle (6.4,-3.9);
\coordinate[align=center,text width=1,label=center:{\parbox{2cm}{\centering Reelay\\(Zenoh)}}] (R) at (5.4,-2.9);


\draw [draw=black, fill=white, very thick] (7.4,-1.9) rectangle (9.4,-2.9);
\coordinate[align=center,text width=1,label=center:{\parbox{2cm}{\centering Carla\\Zenoh Bridge}}] (R) at (8.4,-2.4);

\draw [draw=black, fill=white, very thick] (7.4,-3.1) rectangle (9.4,-4.3);
\coordinate[align=center,text width=1,label=center:{\parbox{2cm}{\centering Carla Simulator}}] (R) at (8.4,-3.7);

\draw [draw=black, fill=white, very thick] (1.4,-1.9) rectangle (3.4,-3.9);
\coordinate[align=center,text width=1,label=center:{\parbox{2cm}{\centering Zenoh\\Router}}] (Router) at (2.4,-2.9);

\draw [draw=black, fill=white, very thick] (10.8,0) rectangle (11.2,-2);
\coordinate[align=center,text width=1,label=center:{\parbox{2cm}{\centering Zenoh-DDS Bridge}}] (Router) at (11,0.5);

\node at (5.0,-0.5) [circle,fill,inner sep=1.5pt]{};
\node at (5.2,-0.5) [circle,fill,inner sep=1.5pt]{};
\node at (5.4,-0.5) [circle,fill,inner sep=1.5pt]{};
\node at (8.4,-0.5) [circle,fill,inner sep=1.5pt]{};
\node at (8.4,-1.5) [circle,fill,inner sep=1.5pt]{};
\node at (11,-0.5) [circle,fill,inner sep=1.5pt]{};
\node at (11,-1.5) [circle,fill,inner sep=1.5pt]{};
\node at (2.4,-1.5) [circle,fill,inner sep=1.5pt]{};
\node at (5.4,-1.5) [circle,fill,inner sep=1.5pt]{};

\draw[very thick] (0,-0.5) -- (12,-0.5);
\coordinate[label=above right:{DDS Network}] (NetDDS) at (0,-0.5);

\draw[very thick] (0,-1.5) -- (12,-1.5);
\coordinate[label=above right:{Zenoh Network}] (NetZenoh) at (0,-1.4);

\draw[rounded corners=8mm,dashed] (3,0)--(9.4,0)--(10.4,-2.5)--(12.4,-2.5)--(12.4,1.2)--(9.4,3.4)--(3,3.4)--cycle;
\coordinate[label=left:{Host 1}] (NetZenoh) at (3,1.6);


\draw[rounded corners=6mm,dashed] (0.5,-1.7)--(10.2,-1.7)--(10.2,-4.5)--(0.5,-4.5)--cycle;
\coordinate[label=left:{Host 2}] (NetZenoh) at (0.5,-3.1);

\end{tikzpicture}
\caption{Container deployment setup distributed over two host machines}
\label{fig:setup2}
\end{figure}
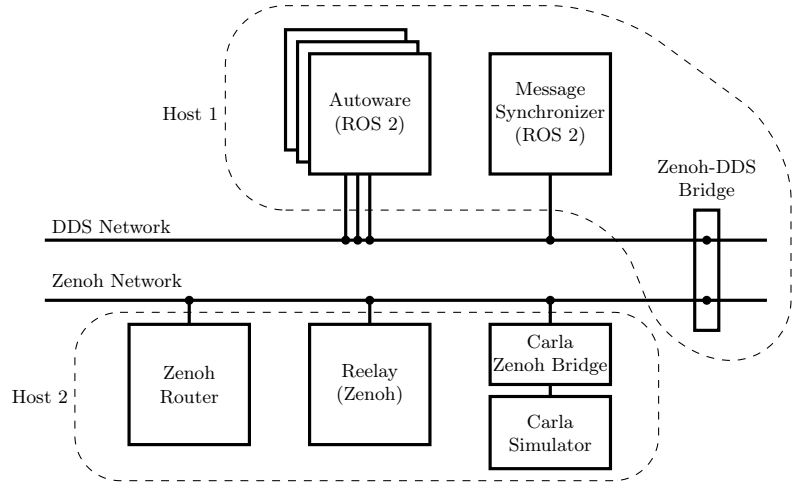

Alternatively, we explore a second deployment setup, illustrated in Figure~\ref{fig:setup2}, where Autoware and the simulation/verification components run on separate host machines. Despite the second setup being more complicated and incurring additional networking penalties, separating the system under test (Autoware) from the testing environment may be beneficial and even necessary if the deployment platform differs from the development. We include the Zenoh-DDS bridge on the first host as Zenoh's user-friendly configuration options allow for granular control over message flow between host machines. We can leverage this to allow or deny specific topics, granting the system architect more control and potentially reducing network bandwidth usage. 

The final piece of these architectural setups is the role of message synchronization, explained in the following section. 

\subsection{Message Synchronizer Participant}
An important challenge in Autoware or similar distributed software stacks is the lack of synchronization across message streams. Since the pub/sub participants operate independently and exchange data asynchronously, ensuring all participants maintain a unified system view is not easy. This is particularly problematic for runtime verification tools such as Reelay, which requires a synchronized input message stream.

To address this challenge, we have developed a ROS message synchronizer participant that can be deployed with other Autoware nodes. This participant employs ROS \texttt{message\_filters} utility library under \texttt{ApproximateTime} policy for synchronization purposes. The \texttt{ApproximateTime} policy does not require messages to have identical timestamps for synchronization, unlike the alternative \texttt{ExactTime} policy. Messages with slightly varying timestamps are chosen to be combined, regardless of their arrival time. This ensures messages subjected to arbitrary networking or processing delays can also be synchronized. The policy finds the best match for a message set based on their timestamps without needing a time difference threshold to define the matching time interval. Each message is used only once, ensuring synchronization sets do not overlap. The sets are contiguous, meaning there is no unused time between consecutive sets, and each set is as small as possible, ensuring tight synchronization and minimal delay.

However, as a limitation, the \texttt{ApproximateTime} policy can synchronize at most nine messages by design. Hence, our synchronizer subscribes to nine different messages from two distinct Autoware packages, generated and published by the Autoware stack, to check the current operational status of the vehicle in simulation. After synchronization, these messages are unified and published under a dedicated runtime verification topic in the network. Table~\ref{tab:message_list} provides a list of selected messages used in this experiment and their corresponding package names with their publishing rates in our deployment setups. 

The message synchronizer's publishing rate heavily depends on the frequency of the messages it subscribes to, which are as low as 3 msg/s as shown in Table~\ref{tab:message_list}. We have not conducted intensive performance tests like those presented in Section~\ref{sec:evaluation} as limited by simulation speeds, yet our observations indicate that the overhead of the Reelay verification tool on communication is negligible at these publishing rates.

\begin{table}[tb]
\centering

\caption{Evaluation of deployment setups}
\small
\begin{tabular}{>{\ttfamily\bfseries}l@{\hskip 0.5cm}>{\ttfamily}l@{\hskip 0.5cm}>{}r@{\hskip 0.5cm}>{}r}
\toprule
\textbf{Package Name} & \textbf{Message} & Setup1 & Setup2\\
\midrule
autoware\_auto\_vehicle\_msgs & SteeringReport & 25.10 & 23.55 \\
autoware\_auto\_vehicle\_msgs & GearReport & 25.02 & 23.69 \\
autoware\_auto\_vehicle\_msgs & ControlModeReport & 24.95 & 22.59 \\
autoware\_auto\_vehicle\_msgs & TurnIndicatorsReport & 24.88 & 22.68 \\
autoware\_auto\_vehicle\_msgs & HazardLightsReport & 25.07 & 23.70 \\
autoware\_auto\_vehicle\_msgs & Engage & 9.95 & 7.95 \\
tier4\_external\_api\_msgs & FailSafeStateStamped & 9.95 & 7.84 \\
tier4\_external\_api\_msgs & Emergency & 10.05 & 8.90 \\
tier4\_external\_api\_msgs & VehicleStatusStamped & 4.99 & 2.97 \\
\bottomrule
\end{tabular}
\label{tab:message_list}
\end{table}

\subsection{Evaluating deployment setups}
In our deployment experiments, we have tested both setups. The first setup achieves a very low simulation speed, as low as 5 simulation frames per second on a machine whose specifications are described in Section~\ref{sec:evaluation}. Although it is still functional at those simulation speeds, it raises concerns regarding the practicality of simulation. This is because simulation-based testing at early stages considerably depends on visualization, where the engineer first visually validates the scenario in the simulation. Such low performance on the simulation side does not help such development workflows.

We have implemented the distributed approach for our second setup by employing an additional machine mirroring the specifications of the initial one. This approach offloads some computationally intensive tasks to the other machine, leading to improved simulation speed and a smoother visual experience for the engineer, as expected. We have executed simulations more smoothly, and the runtime verification engine subscribes to the synchronized topic. Containerization is crucial in ensuring flexibility and modularity within these deployment configurations. Furthermore, it is essential to acknowledge that the process of migrating between these configurations would be significantly more cumbersome without containerization.

\section{Discussion}

In this paper, we presented a containerized solution for runtime verification applications over pub/sub networks. Our benchmark results demonstrated the network performance of standard benchmarks and two deployment setups from a complex real-world example.

One of our main conclusions is that runtime verification containers would be more efficient if they collectively manage tens or even hundreds of properties. This is due to very small messages leading to the under-utilization of the network resources, as shown in our performance experiments. However, too many unrelated properties may lead to higher message synchronization needs with its additional complexity. Therefore, multiple containers to cover a cohesive group of temporal properties seem ideal for complex applications. From real-world examples like Autoware, we know pub/sub networks can go hundreds of participants distributed over several host machines. Therefore, it is not unrealistic that a software-defined vehicle can have multiple runtime verification containers that monitor and verify functional properties at runtime.

In our experiments, we have used push-based subscriptions on the pub/sub-network. We note that the alternative pull-based subscriptions can be more efficient and reliable for runtime verification participants as passive network observers. This gives the the time-synchronization ability to the verification component and pull batches of topic updates at will, further improving network throughput. However, the current support for pull-based subscribers is not extensive and not well-documented for existing pub/sub implementations. We intend to study pull-based subscribers as they develop. 

Binary encoded message formats like \texttt{protobuf} and \texttt{flatbuffers} are also interesting for runtime verification purposes. One advantage of Zenoh compared to DDS implementations is that Zenoh does not dictate a particular wire format. Therefore, applications can use any format as long as both publisher and subscriber know the format. This brings several opportunities to specialize verification participants utilizing certain characteristics of these formats. 

In conclusion, our exploration of containerized runtime verification demonstrates the promise of this approach for efficient and scalable monitoring of publish/subscribe networks. Containerization offers significant advantages for streamlined deployment and management, promoting portability and facilitating adaptation to evolving verification needs. We envision a future of highly flexible and scalable runtime verification solutions using containers within software-defined architectures.

\printbibliography

\end{document}